\documentclass[iicol, pdflatex, sn-basic]{sn-jnl}

\usepackage{lineno}
\usepackage{graphicx}
\usepackage{multirow}%
\usepackage{amsmath,amssymb,amsfonts}%
\usepackage{amsthm}%
\usepackage{mathrsfs}%
\usepackage[title]{appendix}%
\usepackage{xcolor}%
\usepackage{textcomp}%
\usepackage{manyfoot}%
\usepackage{booktabs}%
\usepackage{algorithm}%
\usepackage{algorithmicx}%
\usepackage{algpseudocode}%
\usepackage{listings}%

\begin{document}

\title[Rayleigh and Interstellar Polarization for Evolved Late-Type Stars]{Modeling of Rayleigh Scattering and Interstellar Polarization for Evolved Late-Type Stars}

%%=============================================================%%
%% Prefix	-> \pfx{Dr}
%% GivenName	-> \fnm{Joergen W.}
%% Particle	-> \spfx{van der} -> surname prefix
%% FamilyName	-> \sur{Ploeg}
%% Suffix	-> \sfx{IV}
%% NatureName	-> \tanm{Poet Laureate} -> Title after name
%% Degrees	-> \dgr{MSc, PhD}
%% \author*[1,2]{\pfx{Dr} \fnm{Joergen W.} \spfx{van der} \sur{Ploeg} \sfx{IV} \tanm{Poet Laureate} 
%%                 \dgr{MSc, PhD}}\email{iauthor@gmail.com}
%%=============================================================%%

\author*[1]{\fnm{Richard} \sur{Ignace}}\email{ignace@etsu.edu}

\author[2]{\fnm{Christiana} \sur{Erba}}\email{christi.erba@gmail.com}

\author[3]{\fnm{Kaylee} \sur{DeGennaro}}\email{kaylee\_degennaro@brown.edu}

\author[1]{\fnm{Gary D.} \sur{Henson}}\email{hensong@etsu.edu}

\affil*[1]{\orgdiv{Department of Physics \& Astronomy}, \orgname{East Tennessee State University}, \orgaddress{\street{}, \city{Johnson City}, \postcode{37614}, \state{TN}, \country{USA}}}

\affil*[2]{\orgname{Space Telescope Science InstituteBrown University}, \orgaddress{\street{3700 San Martin Drive}, \city{Baltimore}, \postcode{21218}, \state{MD}, \country{USA}}}

\affil*[3]{\orgdiv{Department of Physics}, \orgname{Brown University}, \orgaddress{\street{182 Hope St.}, \city{Providence}, \postcode{02912}, \state{RI}, \country{USA}}}

%%==================================%%
%% sample for unstructured abstract %%
%%==================================%%

\abstract{
Evolved late-type stars are frequently identified as photometric and spectroscopic variables, such as Mira-type or semi-regular variable objects. These stars can also be polarimetrically variable, an indicator of non-spherical geometry for spatially unresolved sources. Departures from symmetry can arise in a number of ways, such as the presence of a binary companion (e.g., multiple illumination sources for scattered light), brightness variations in the stellar atmosphere (e.g., large convective cells), or aspherical circumstellar envelopes (e.g., disks or aspherical stellar winds). Common polarigenic opacities for cool stars include Rayleigh scattering and dust scattering. The classic wavelength dependence of $\lambda^{-4}$ for Rayleigh single scattering is generally straightforward; however, that signature can be confounded by interstellar polarization (ISP).  
We explore strategies for interpreting polarimetric observations when the interstellar polarization (ISP) cannot be removed. We introduce a ``hybrid'' spectrum that includes both Rayleigh polarization for a stellar contribution and the Serkowski Law for an interstellar contribution. We find the polarization spectral slope can be more shallow than expected from Rayleigh scattering alone. For stellar variability, shorter wavelengths give higher amplitude changes when Rayleigh scattering dominates the interstellar signal. Quite anomalous slopes can occur over limited wavelength intervals if the stellar and interstellar position angles differ by $90^\circ$. Results of the models are discussed in the context of photopolarimetry methods, and an application is considered in terms of variable polarization from the carbon star, R~Scl.
}

%%================================%%
%% Sample for structured abstract %%
%%================================%%

% \abstract{\textbf{Purpose:} The abstract serves both as a general introduction to the topic and as a brief, non-technical summary of the main results and their implications. The abstract must not include subheadings (unless expressly permitted in the journal's Instructions to Authors), equations or citations. As a guide the abstract should not exceed 200 words. Most journals do not set a hard limit however authors are advised to check the author instructions for the journal they are submitting to.
% 


%%\pacs[JEL Classification]{D8, H51}

%%\pacs[MSC Classification]{35A01, 65L10, 65L12, 65L20, 65L70}

\maketitle

\section{Introduction}\label{Intro}

Except for a limited number of cases in which stars
can be resolved through imaging \citep[e.g.,][]{1999AJ....117..521V}, stars are so distant as to be effectively point sources. Spatial information about their atmospheres and circumstellar environments are often inferred from spectral diagnostics derived from their spectral energy distributions, from tracing the Doppler effect in spectral lines, or through studies of time variability. Another valuable tool for discerning spatial information about unresolved stars is polarimetry \citep[e.g.,][]{2010stpo.book.....C}, since a star that is spherically symmetric will display no net linear polarization, even though the emergent intensities along individual
rays through the source are polarized \citep{1960ratr.book.....C}. The reason is that linear
polarization is subject to geometrical cancellation \citep[e.g.,][]{1977A&A....57..141B};  consequently,
a net polarization measured in an unresolved star is indicative of a geometry that is not spherical.

%However, there is a challenging aspect to this situation, namely
The polarizing influence of the interstellar medium (ISM) through which the starlight must travel presents a challenge to this interpretation. Even if a star is
spherically symmetric and intrinsically unpolarized, an interstellar polarization (ISP), arising from the interaction of the starlight with the ISM, will generally be present \citep[e.g.,][]{2025ApJS..276...15P}. ISP values are often approximated well by the Serkowski Law \citep{1975ApJ...196..261S}, which involves a few free parameters \citep[e.g.,][]{1992ApJ...386..562W}. An ensemble of strictly unpolarized stars at different distances and directions display a range of polarization values \citep[e.g.,][]{2025ApJS..276...15P}.

Importantly, even when the ISP is not removed, it is still possible to infer some properties of the stellar contribution. The ISP is expected to be smooth with wavelength, have constant polarization position angle (PA), and  not vary with time. Thus, polarization measurements that deviate from the Serkowski Law, that show PA rotation with wavelength, or that exhibit time variability are all signatures of the stellar contribution.  \cite{2025Ap&SS.370...57I} have explored how to infer stellar effects when the ISP contribution has not or cannot be removed in the case of hot stars. For those objects, gray Thomson scattering by free electrons is often important, and asymmetries in the circumstellar matter can produce spectral signatures of an intrinsic stellar contribution (such as the emission line ``dilution effect''). By contrast, the coolest stars of classes K (non-coronal) and M have neutral Hydrogen, and their spectra contain only low ionization states of trace metal species \citep[e.g.,][]{2009ssc..book.....G}. For these cases, the polarigenic opacities are expected to be dominated by dust and/or Rayleigh scattering by atomic and molecular species \citep[e.g.,][]{2010stpo.book.....C}.

Rayleigh scattering is interesting for its well-known wavelength dependence of $\lambda^{-4}$ through the optical and infrared (IR) regimes.  For thin Rayleigh scattering in circumstellar envelopes, this chromatic signature is expected of the stellar polarization contribution.  This characteristic spectral signature has been repeatedly observed in cool stars \citep[e.g.,][]{1979AJ.....84.1200C, 1991AJ....101.1735J, 2001AJ....122.2017S, 2006ApJ...639.1053B, 2016A&A...591A.119A, 2023A&A...677A..96N}.  

Relatively recent stellar atmosphere modeling that includes Rayleigh scattering indicates that limb polarization can be significant in evolved cool stars (whereas it is small in dwarf stars at the same spectral type; \citealt{2015A&A...575A..89K, 2016A&A...586A..87K}). Consequently, when asymmetries are present in the atmosphere itself, such as the purportedly convective ``hot spots'' observed in Betelgeuse \citep{2016A&A...588A.130M,2018A&A...609A..67K}, polarization signatures may become detectable. However, the chromatic signature may not follow Rayleigh's law for the optical owing to the complexities of radiative transfer in stellar atmospheres \citep{1969ApL.....3..165H, 1986ApJ...307..261D}.

Since it is not always easy to remove the ISP contribution, the focus of this paper is to model ``hybrid'' scenarios, where the Serkowski law and Rayleigh scattering are combined to produce synthetic linearly polarized continua, and to examine how such hybrid models deviate from the expected $\lambda^{-4}$ trend through a parameter study. In particular, when the resulting trend is much more shallow than that expected from Rayleigh scattering, it is tempting to conclude that the polarigenic opacity may instead arise from dust or Mie scattering \citep[e.g.,][]{1982MNRAS.200...91S}. We thus explore whether polarimetric variability in hybrid models with Rayleigh scattering leads to predictions that can discriminate whether dust scattering is to be favored.  The hybrid models are defined in Section~\ref{model} along with presentation of a parameter study. A discussion of the connections to interpreting observational data appears in Section~\ref{disc}. Summary remarks are given in Section~\ref{summ}. Finally, a special case for the wavelength-dependent continuum polarization slope in the hybrid scenario is described in Appendix~\ref{sec:app}.

\section{The Stellar + Interstellar Hybrid Model}\label{model}

Our hybrid models consist of combining the Serkowski law for the ISP with a Rayleigh scattering signature describing the stellar contribution. We employ the standard I, Q, U, and V Stokes parameters \citep{1960ratr.book.....C}. Stokes I is the total intensity (or flux or luminosity).  Stokes V gives the circular polarization, which is a diagnostic of magnetism and which will be ignored hereafter (so $V=0$). The Stokes Q and U parameters describe the linear polarization. They act as ``vector components'' of the total polarization and thus serve to define the polarization position angle.

In the models that follow, we use the traditional normalized parameters for linear polarization with $q=Q/I$ and $u=U/I$, where the capitalized parameters would have physical units like specific flux or specific luminosity, and the lower-case parameters are fractional \citep{2010stpo.book.....C}. The polarization $p$ and corresponding position angle $\psi$ are thus given by
\begin{equation}
p = \sqrt{q^2+u^2},
\end{equation}
\noindent and 
\begin{equation}
   \tan 2\psi = u/q. 
\end{equation}
\noindent Importantly, the hybrid scenario involves two polarimetric contributions. The imprint of the ISP, when the ISP is small as expected, is a linear combination in the $q$ and $u$ parameters. Denoting the stellar contribution as $q_\ast$ and $u_\ast$ (with corresponding $p_\ast$ and $\psi_\ast$), and the ISP as $q_I$ and $u_I$ (likewise with $p_I$ and $\psi_I$), the resultant total Stokes parameters and associated polarization and PA will be
\begin{eqnarray}
q_{\rm tot} & = & q_\ast+q_I, \label{eq:qtot} \\
u_{\rm tot} & = & u_\ast+u_I,\\
p_{\rm tot} & = & \sqrt{q^2_{\rm tot}+u^2_{\rm tot}},~{\rm and}\\
\tan 2 \psi_{\rm tot} & = & u_{\rm tot}/q_{\rm tot}. \label{eq:psitot}
\end{eqnarray}
\noindent The formulation of linear polarization is vector-like, treating $q$ and $u$ like cartesian coordinates and $p$ and $\psi$ like polar coordinates. The hybrid scenario with both stellar and interstellar contributions is linear in $q$ and $u$, making both $p$ and $\psi$ nonlinear in combination. It is this non-linearity that leads to a variety of spectropolarimetric outcomes, as well as trends that are explored in our parameter study.

\subsection{Rayleigh scattering for the intrinsic polarization of the star}

Polarization signals from late-type stellar sources may arise from asymmetries in the stellar atmosphere, the circumstellar environment, or from a combination of both of these areas. Additionally, physical models of late-type stellar polarimetry typically account for the detailed opacities of molecular species. Since we are not concerned here with this kind of physical model for explaining the sources of asymmetry (which can differ between stars), we instead develop a simple approach that parametrizes the stellar polarization level (regardless of cause). We focus on the combination of the stellar signal with the ISP to clarify what may be robustly concluded from observables.

We briefly review the classical derivation for Rayleigh scattering employed in our parameterization \citep[e.g.,][]{2004RL-book}. For a bound electron that is treated as a damped harmonic oscillator, the cross-section has a Lorentzian shape given by
\begin{equation}
\sigma(\lambda) = \sigma_T\,\left[\left(1-\frac{\lambda^2}{\lambda_0^2}\right)^2+\kappa^2\,\lambda^2\right]^{-1},
\end{equation}
\noindent where $\sigma_T$ is the Thomson cross-section, $\lambda_0$ is the resonance wavelength of the bound electron, and $\kappa$ is related to the damping constant and has units of inverse wavelength. For long wavelengths $\lambda \gg \lambda_0$, the regime of Rayleigh scattering is recovered with $\sigma \propto \lambda^{-4}$. At short wavelengths with $\lambda\ll \lambda_0$, the cross-section becomes constant with $\sigma \approx \sigma_T$.  Near resonance, the cross-section can be large, with $\sigma \approx \sigma_T/(\kappa\lambda_0)^2.$

The condition of resonance is set far in the ultraviolet (UV), whereas our treatment is concerned mainly with optical and near-Infrared (NIR) wavelengths that are accessible from the ground, for which the majority of data have been obtained in linear polarimetry \citep[e.g., the compilation of][]{2001AJ....122.2017S}.  Consequently, the details of resonance are not of interest to the modeling, and for simplicity, we take $\kappa^2 = 2/\lambda_0^2$, which leads to a simplified form of the cross-section as
\begin{equation}
    \sigma(\lambda) = \sigma_T\,\left[1+\left(\frac{\lambda}{\lambda_0}\right)^4\right]^{-1}.
    \label{eq:sigma}
\end{equation}
\noindent Typically, the cross-section of resonance scattering is orders of magnitude larger than that given by pure Thomson scattering, suggesting that $\kappa\lambda_0$ is small. Moreover, the width of that peak is likewise related to $\kappa\lambda_0$. Consequently, 
our simplification is valid in the sense that it produces the standard Rayleigh scattering power-law for optical wavelengths. The form of equation~(\ref{eq:sigma}) essentially ignores the resonance condition and has the form of a short-wavelength plateau and a long-wavelength $\lambda^{-4}$ power law.

Applying the conditions above, we take the stellar polarization to be
\begin{equation}
    p_\ast(\lambda) = p_0\,\left[1+\left(\frac{\lambda}{\lambda_0}\right)^4\right]^{-1}.
    \label{eq:starpol}
\end{equation}
\noindent This form is not tied to any particular physical model of the star or its circumstellar envelope. As long as the scattering is effectively optically thin, the polarization will follow the Rayleigh power law at long wavelengths regardless of whether symmetry breaking is in the atmosphere \citep[e.g.,][]{1986ApJ...307..261D} or the circumstellar envelope.  

With this definition, the free parameters of the stellar polarization are $p_0$, $\lambda_0$, and $\psi_\ast$. For our proof-of-concept study, we choose $\lambda_0=2000~\AA$, then choose $p_0$ to produce a level of polarization at optical wavelengths similar to that observed. For example, $p=1\%$ at $\lambda=5000~\AA$ requires $p_0=40\%$.  This is not to imply that the polarization at UV wavelengths will actually be that high; $p_0$ should be interpreted merely as a ``knob'' that is used to set a polarization level at wavelengths accessible to ground-based facilities.

Finally, the PA for the star, $\psi_\ast$, is a free parameter. The Stokes parameters for the stellar contribution from the model are determined by
\begin{eqnarray}
q_\ast(\lambda) & = & p_\ast\,\cos 2 \psi_\ast,~{\rm and} \\
u_\ast(\lambda) & = & p_\ast\,\sin 2 \psi_\ast.
\end{eqnarray}
\noindent Note that $\psi_\ast$ is a constant with no chromatic variation.  Consequently, with $p_\ast$ varying as $\lambda^{-4}$, both $q_\ast$ and $u_\ast$ are similarly varied.

\subsection{Inclusion of interstellar polarization}

As noted, when starlight is unpolarized, its measurement at Earth will generally have some level of linear polarization arising from its passage through the interstellar medium \citep{1996cduc.conf..155W}. This imprint of ISP is well described by the Serkowski Law \citep{1975ApJ...196..261S} with the following functional form: 
\begin{equation}
    p_I(\lambda) = p_{\rm max}\, e^{-K\,[\ln (\lambda_{\rm max}/\lambda)]^2 },
\end{equation}
\noindent where $p_{\rm max}$, $K$, and $\lambda_{\rm max}$ are parameters specific to the medium through which the starlight travels. The position angle specified by $\psi_I$ is not wavelength-dependent. Note that $K$ is typically of order unity, whereas $\lambda_{\rm max}$, the wavelength at which the maximum ISP $p_{\rm max}$ is achieved, typically falls between $4000-7000$~\AA\ \citep{1992ApJ...386..562W}.

We note that studies indicate interstellar polarization that deviates from the analytic form of the Serkowski Law. Combining UV, optical, and IR data, \cite{1999ApJ...510..905M} found in particular at wavelengths longer than about 1 micron, the interstellar polarization becomes power-law in form rather than Gaussian. Those authors report that the interstellar polarization is consistent with declining as $\lambda^{-2}$ toward the IR. Although our approach can accommodate any wavelength-dependent form for interstellar polarization, we adopt the standard Serkowski Law given that our focus is mainly optical wavelengths and extension toward the UV and the NIR.

For our purposes, we fix $\lambda_{\rm max}$ at 5500~\AA. Using $K=1.68/\lambda_{\rm max}~(\mu {\rm m})$ then results in $K=0.9$ \citep{1992ApJ...386..562W}. Our parameter study allows for $p_{\rm max}$ and $\psi_I$ to vary. The Stokes parameters thus become
\begin{eqnarray}
q_I(\lambda) & = & p_I\,\cos 2 \psi_I,~{\rm and} \\
u_I(\lambda) & = & p_I\,\sin 2 \psi_I,
\end{eqnarray}
\noindent with functional forms that are similar to those applied for the stellar contribution. Since the ISP polarization $p_I$ follows the Serkowski Law in its wavelength dependence, the terms for $q_I$ and $u_I$ follow similarly, since $\psi_I$ has no wavelength dependence.

\begin{figure*}[h]%
\centering
\includegraphics[width=\textwidth]{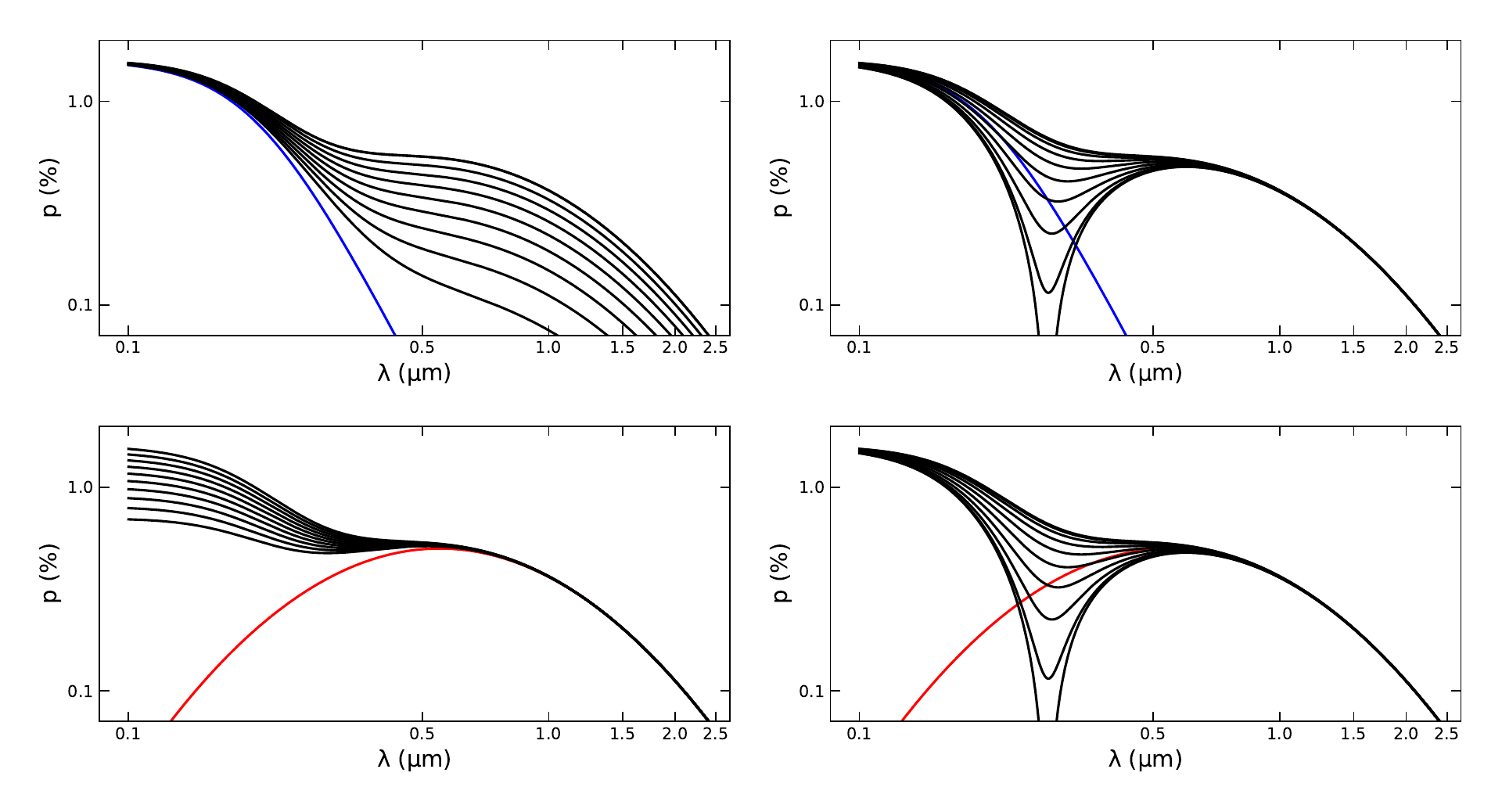}
\caption{Models of continuum polarization spanning from FUV through NIR wavelengths, displayed in log-log plots for parameters identified in Tab.~\ref{tab1}. In the two left panels, the position angles $\psi_\ast$ and $\psi_I$ are held constant while $p_{\rm max}$ is varied (upper left) or $p_\ast$ is varied (lower left).  The blue curve at upper left is the fixed stellar polarization whereas the red curve at lower left is the fixed interstellar polarization.
In the two right panels, the polarization scales $p_{\rm max}$ and $p_\ast$ are held constant. At upper right the stellar position angle $\psi_\ast$ varies, whereas at lower right it is $\psi_I$ that varies.  Again blue and red are for the fixed stellar and interstellar polarizations, respectively.  Also, because $p_{\rm max}$ and $p_\ast$ are fixed, the upper and lower right panels are redundant, since what actually matters is the relative position angle, $|\psi_\ast-\psi_I|$.}\label{fig1}
\end{figure*}

\begin{table}
\caption{Model Parameters$^a$ for Figures.  \label{tab1} }
\begin{tabular}{lllll}
\hline\hline Panel & $p_{\rm max}$ & $\psi_I$ & $p_\ast$ & $\psi_\ast$ \\ 
 & (\%) & $(^\circ)$ & (\%) & $(^\circ)$ \\
 \hline
 Upper left & 0.05--0.50$^b$ & 0 & 1.6 & 0 \\
 Lower left & 0.5 & 0 & 0.7--1.6$^c$ & 0 \\
 Upper right & 0.5 & 0 & 1.6 & 0--90$^d$ \\
 Lower right & 0.5 &  0--90$^d$ & 1.6 & 0 \\ \hline
\end{tabular}

$^a$ In all cases $\lambda_{\rm max}=0.55~\mu$m and $K=0.9$. \\
$^b$ Steps of 0.05. \\
$^c$ Steps of 0.1. \\
$^d$ Steps of 10. 
\end{table}

\begin{figure}[h]%
\centering
\includegraphics[width=\columnwidth]{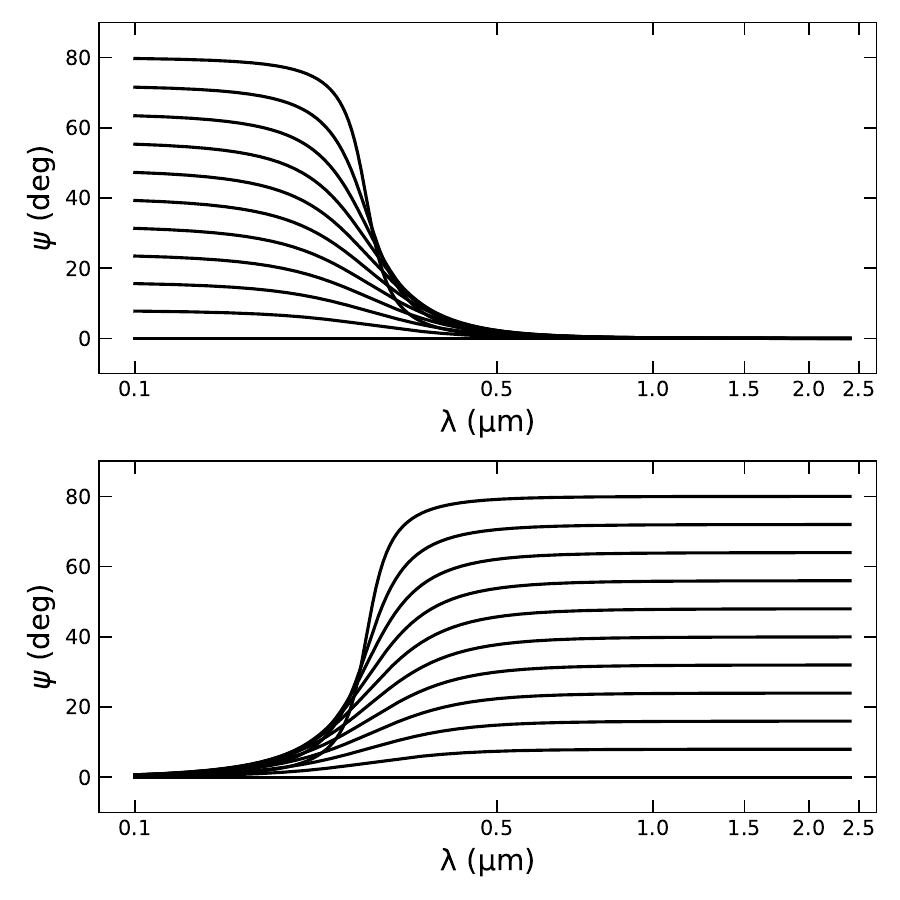}
\caption{Polarization position angles plotted with wavelength (c.f., the right-side panels for Fig.~\ref{fig1} and Tab.~\ref{tab1}).  In the upper panel, the fact that $\psi_I$ is constant is signified by all PAs converging to zero at longer wavelengths. In the lower panel, the trend is opposite, where all PAs converge to zero at short wavelengths because $\psi_\ast$ is now  held constant.}\label{fig2}
\end{figure}

\subsection{Parameter Study}

Our hybrid model combines the stellar and interstellar contributions by summing the Stokes $q$ and $u$ parameters and producing the total polarization and net PA following equations~(\ref{eq:qtot})--(\ref{eq:psitot}). Since $\psi_\ast$ and $\psi_I$ are uncorrelated, the chromatic dependence of the resultant polarization now generally deviates from the trends expected from either simple Rayleigh scattering or the Serkowski law. Additionally, the net position angle is not generally constant with wavelength.  

\begin{figure*}[h]%
\centering
\includegraphics[width=0.90\textwidth]{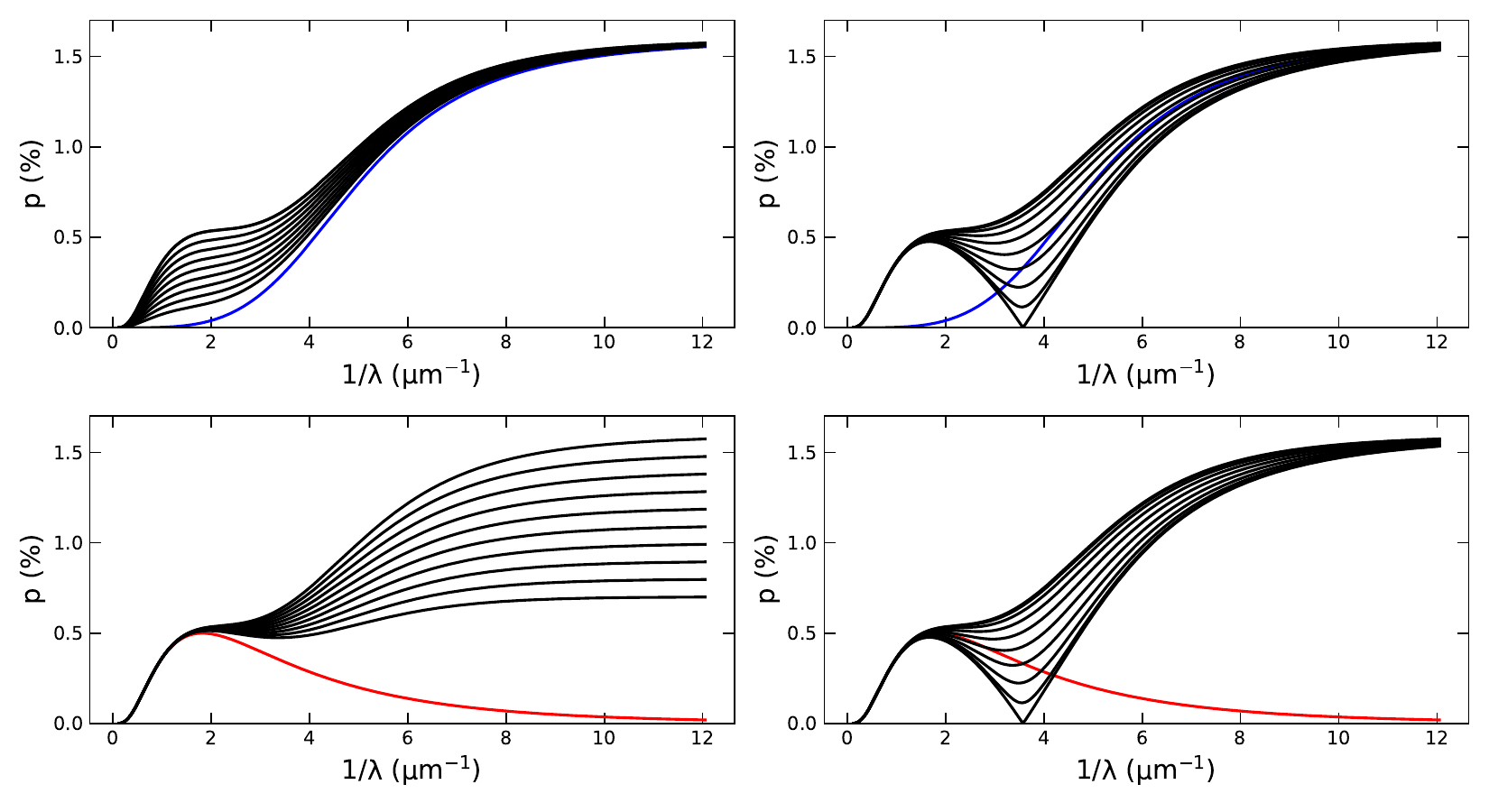}
\caption{The same models as in Fig.~\ref{fig1} now plotted with polarization as linear against inverse wavelength, $1/\lambda$, with $\lambda$ in microns. This version portrays the models in a format sometimes adopted by observers.}\label{fig3}
\end{figure*}

Importantly, the hybrid regime typically falls in the optical band, where the Serkowski Law peaks, as opposed to the peak from Rayleigh scattering located in the UV regime. To illustrate this, Figures~\ref{fig1} and \ref{fig2} show the primary results of our parameter study. Figure~\ref{fig1} displays spectral energy distributions in polarized light. Each panel shows the variation of the model polarization (which includes both stellar and interstellar contributions) with wavelength in log-log plots extending from the UV, through the optical, and into the NIR regime. Model parameters for each panel are given in Table~\ref{tab1}. The blue curves indicate the trend resulting from Rayleigh scattering alone; the red curves indicate the trend resulting from the Serkowski law alone. The upper panel at left shows calculations when Rayleigh scattering is fixed and $p_{\rm max}$ is varied. In this case, note that $\psi = \psi_I=\psi_\ast$. The lower panel at left shows the result for a fixed Serkowski law, but with $p_0$ allowed to increase while keeping the polarization PAs aligned.

The right side of Figure~\ref{fig1} shows what happens when $p_{\rm max}$ and $p_0$ are fixed, but $\psi_I$ and $\psi_\ast$ are allowed to rotate with respect to each other. The upper panel shows the result when $\psi_I$ rotates with respect to $\psi_\ast$, and the lower panel illustrates the opposite case. The figure demonstrates the resultant degeneracy, since what matters for the polarization is the relative difference in the stellar and interstellar PAs.  However, the effective PA, $\psi_{\rm tot}$ becomes chromatic, as illustrated in Figure~\ref{fig2}, with upper and lower panels corresponding to the models on the right side of Figure~\ref{fig1}.

Two key results immediately follow: 
first, the polarization for the hybrid scenario is no longer a power-law with the $-4$ exponent expected from Rayleigh scattering. In fact, {\it the curves are not power-laws at all}, a result that is specifically highlighted for an analytic test case in Appendix~\ref{sec:app}. As a result, a range of possibilities for the predicted spectropolarimetric shapes are produced. When the PA difference between $\psi_I$ and $\psi_\ast$ is far from $90^\circ$, the polarization approximately follows a power-law trend {\it in segments}, but in isolation, the polarization is generally less steep than expected for Rayleigh scattering.

However, when the respective PAs get close to $90^\circ$, the steepness from Rayleigh scattering ensures the polarization will drop to zero at some wavelength if $p_0>p_{\rm max}$. When this happens, slopes that are quite anomalous from the canonical Rayleigh result can be produced over limited wavelength intervals. Indeed, even positive slopes can occur over intervals in wavelength, and negative slopes could be much steeper than expected from Rayleigh.
In practice, such effects may be difficult to measure, since the polarization becomes small. Additionally, as highlighted in Section~\ref{disc}, many of the available datasets are based on photopolarimetry, which may suppress the effects seen in the models through flux-averaging over wavelength for broad passbands.

Given that the Serkowski law is relatively flat in the optical, a crude estimate for the critical wavelength, $\lambda_{\rm c}$, where the polarization drops to zero is
\begin{equation}
    \lambda_{\rm c} \sim \left(\frac{p_0}{p_{\rm max}}\right)^{1/4}\,\lambda_0.
\end{equation}
\noindent The actual solution for $\lambda_{\rm c}$ is not analytic and will depend on $K$ and $\lambda_{\rm max}$. The consequence of this behavior, should it occur, is a strong deviation from a power-law trend with the possibility of a steepenig negative slope shortward of $\lambda_{\rm c}$ followed by a positive slope longward of it.

For polarimetric studies of cool evolved stars, polarimetric measurements are sometimes displayed in plots against inverse wavelength, $1/\lambda$.  This can be useful for visually assessing whether polarization rises toward shorter wavelengths and if so, how steeply.  As a convenience, the model results from Figure~\ref{fig1} are replotted in Figure~\ref{fig3} using this alternative format.  

\section{Discussion}\label{disc}

\subsection{Photopolarimetry-based studies}

Since observed stellar linear polarization values tend to be small, studies often use color-dependent photopolarimetry to increase signal-to-noise.  For quite cool stars like the giants and supergiants of M spectral class, photopolarimetry at optical and NIR wavelengths will be on the Wien side of the flux distribution. The method of conducting photopolarization involves measuring source fluxes in Stokes Q and Stokes U, and then measuring the Stokes- I flux. From these, the relative polarizations $q$ and $u$ are formed, and if desired $p_{\rm tot}$ and $\psi_{\rm tot}$ can be calculated.

For modeling purposes, we consider a photometric passband ``X'' with characteristic wavelength $\lambda_X$, appropriate for a flat spectrum as a flux-weighted effective wavelength.  The passband Stokes-I flux is given by

\begin{equation}
    f_{I,x} = \int f_\nu(\lambda) \, T_X(\lambda) \, d\lambda,
\end{equation}

\noindent where $T_X$ is the transmission of the filter such that

\begin{equation}
    \int T_X(\lambda)\, d\lambda = 1.
\end{equation}

\noindent The linear polarization derives from measures of the Stokes Q and U passband fluxes, with

\begin{eqnarray}
   f_{Q,X}&=&\int f_\nu(\lambda) \, T_X(\lambda) \,p_{\rm tot}(\lambda)\, \cos 2\psi_{\rm tot}(\lambda)\,d\lambda,~~~~{\rm } \\
   f_{U,X}&=&\int f_\nu(\lambda) \, T_X(\lambda) \,p_{\rm tot}(\lambda)\, \sin 2 \psi_{\rm tot}(\lambda)\,d\lambda,~~~~~  
\end{eqnarray}

\noindent where $p_{\rm tot}$ and $\psi_{\rm tot}$ are the total polarization and corresponding position angle.  We introduce the polarized flux as $f_{\rm p} = \sqrt{f_{Q}^2 + f_{U}^2}$, and the relative polarization will be $p = f_{\rm p}/f_\nu$.

Thus photopolarimetry making use of broad passbands typically assumes the polarization measures are represented approximately by the quoted effective wavelengths of the passbands.  However, if the spectra are quite steep over the span of wavelengths sampled by the passbands, the quoted effective wavelengths may not be representative of the polarization sampling.

\begin{figure*}[h]%
\centering
\includegraphics[width=0.90\textwidth]{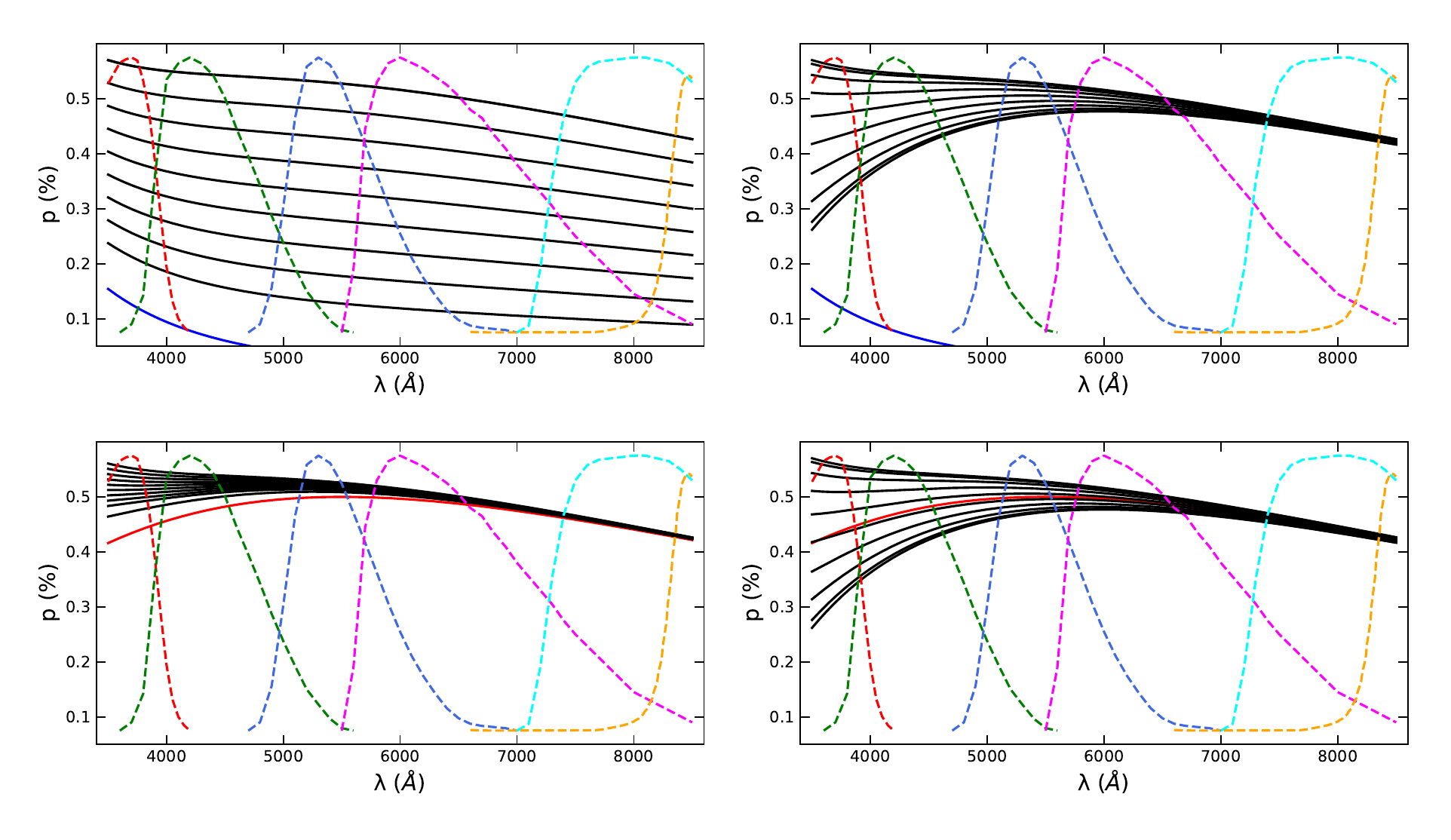}
\caption{The calculations from Fig.~\ref{fig1} for polarization now limited to optical wavelengths. Superposed in colored dotted curves are the standard passband responses for UBVRIZ filters as red, green, blue, magenta, cyan, and orange. These are peak normalized and scaled to the figure.}\label{fig4}
\end{figure*}

\begin{figure}[h]%
\centering
\includegraphics[width=\columnwidth]{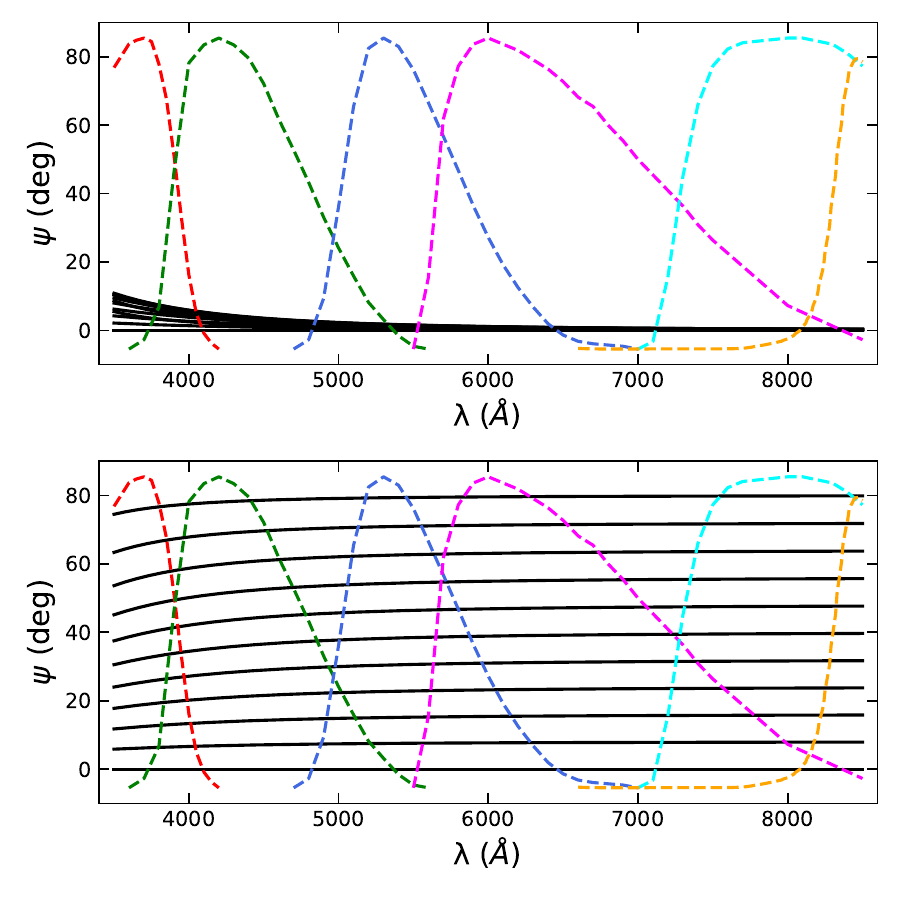}
\caption{The calculations from Fig.~\ref{fig2} for polarization position angle now limited to optical wavelengths. Superposed in colored dotted curves are the standard passband responses for UBVRIZ filters. These are peak normalized and scaled to the figure.}\label{fig5}
\end{figure}

Figures~\ref{fig4} and \ref{fig5} show the same model curves of Figures~\ref{fig1} and \ref{fig2}, respectively, but now limited to the optical wavelengths.  Overplotted in color are the filter responses for UBVRIZ passbands\footnote{The UBVRIZ notation shorthands two standard photometric systems typically used in optical photometry: the UBVRI system \citep[e.g.,][]{Bessell1990,Bessell2005} and the {\it ugriz} system \citep[e.g.,][]{Smith2002}.}.  These are peak normalized to fit the scale of the figures for reference.  These provide examples for how different passbands would sample different regimes of polarization and position angle that are governed mainly by Rayleigh or Serkowski trends, or by a mix, for the chosen parameters of these examples. Note that the polarizations in Figure~\ref{fig1} were plotted logarithmically but are linear in Figures \ref{fig4} and \ref{fig5}.

%\begin{figure}[h]%
%\centering
%\includegraphics[width=\columnwidth]{PAv1ovlambda_UBVRIZ.pdf}
%\caption{New version of  crossPA.pdf. UBVRIZ bands are reversed due to 1/$\lambda$ scale. Domain is 3500-8500 angstroms, or 0.35-0.85 microns.}\label{fig6}
%\end{figure}

\subsection{The carbon star, R~Scl}

Being on the Wien side of the spectral distribution for wide photometric bands suggests that there could be color-term effects. In the case of cool stars, the properties of the photometry may  be biased toward the red side of those responses. Adding the steep response of Rayleigh scattering, which is biased toward the blue side of the passband, can compensate for this effect in the polarized flux. 

The color terms can be expressed by flux-weighted wavelengths.  We introduce two effective wavelengths, for the flux spectrum (with subscript ``f'') and for the polarized flux spectrum (with subscript ``p''), as

\begin{eqnarray}
    \lambda_{\rm f ,X}&=&\frac{\int \lambda\, f_\nu(\lambda) \, T_X(\lambda) \,d\lambda}{\int f_\nu(\lambda) \, T_X(\lambda) \,d\lambda},~{\rm and} \\
    \lambda_{\rm p,X}&=&\frac{\int \lambda\, f_{\rm p}(\lambda) \, T_X(\lambda) \,d\lambda}{\int f_{\rm p}(\lambda) \, T_X(\lambda) \,d\lambda}.
\end{eqnarray}

\noindent On the Wien side of the stellar flux distribution, which is appropriate for quite cool stars, the change in effective wavelength for the passband $\lambda_{\rm f ,X}$ can shift significantly longward of $\lambda_X$.  For optically thin circumstellar Rayleigh scattering, one expects $f_{\rm p} \propto f_\nu / \lambda^4$.  

\begin{figure}[h]%
\centering
\includegraphics[width=\columnwidth]{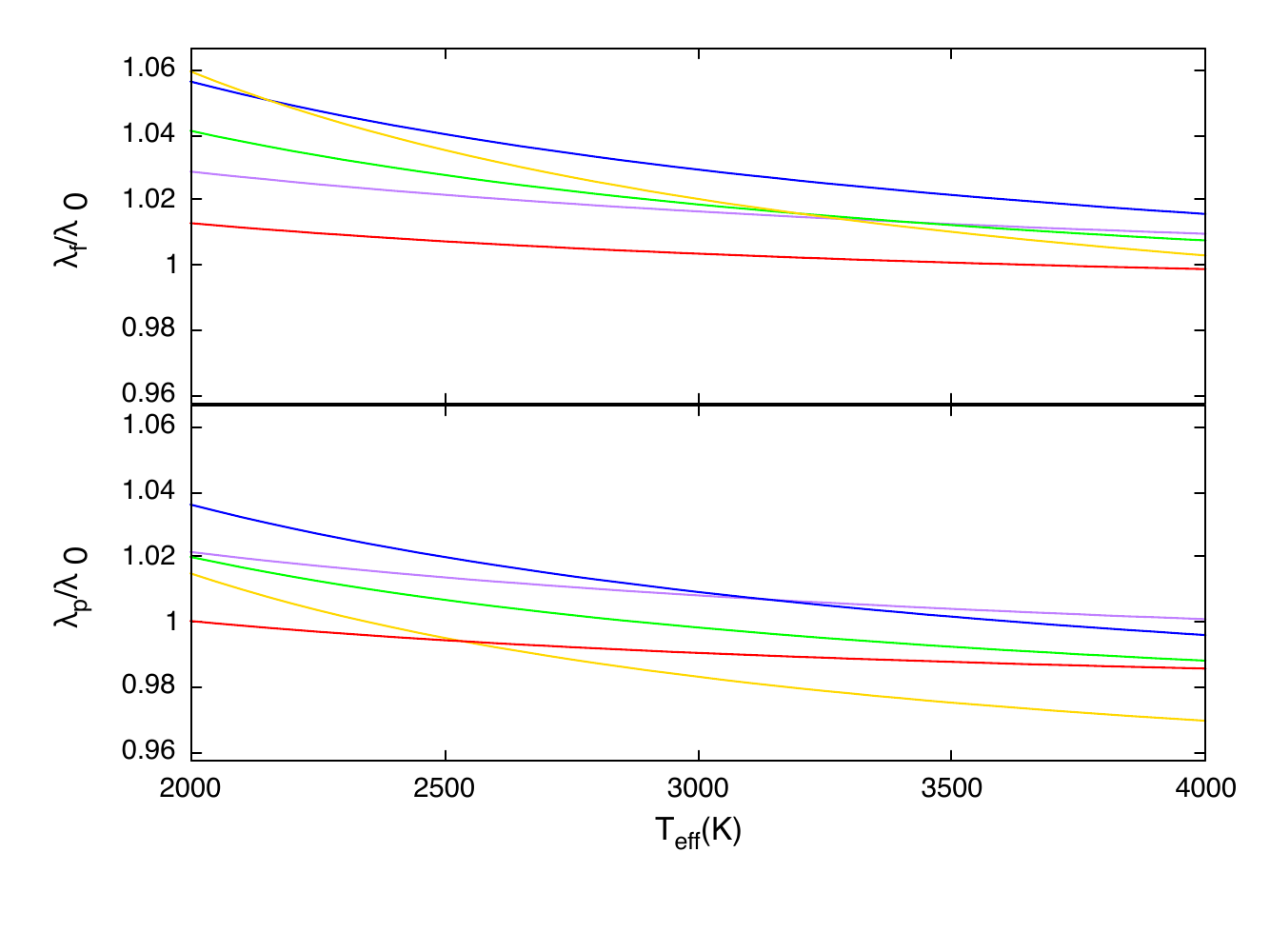}
\caption{Calculations for the flux-weighted effective wavelengths for the total flux spectrum $\lambda_{\rm f}$ (top) and the polarized spectrum $\lambda_{\rm p}$ (bottom) as a function of effective temperature using a blackbody spectrum. These are expressed as ratios relative to the respective effective wavelengths $\lambda_{0}$ described in text.  The colors are for bands U (purple), B (blue), V (green), R (yellow), and I (red).}\label{fig6}
\end{figure}

We apply a blackbody for the flux spectrum and adopt the Rayleigh scattering limit for a circumstellar scattering polarization. The resulting calculations for the flux-weighted wavelengths are shown in Figure~\ref{fig6}.  The upper panel is for $\lambda_{\rm f}$; the lower panel is for $\lambda_{\rm p}$. These are normalized to $\lambda_{0}$, that we introduce as the flux-weighted wavelength assuming a flat spectrum. The curves of different colors represent the set of bands ``$X$'' for UBVRI with purple, blue, green, yellow, and red, respectively. The calculations are only for cool effective temperatures from 2000~K to 4000~K.

Note that for $\lambda_{\rm f}$, all the wavelength shifts are positive.  The Wien peak is longward of 7300~\AA\ for all these temperatures. By contrast, $\lambda_{\rm p}$ can be negative in some cases at the hotter end of the models.  This implies effective wavelengths that have shifted shortward of $\lambda_{0}$, and arises from the additional weighting by $\lambda^{-4}$ owing to Rayleigh scattering.

The shifts are relatively small -- at the level of a few percent at the coolest temperatures -- but when looking for evidence of Rayleigh scattering from the polarized spectra, a few percent correction is magnified by the 4th power. If all filters had the same correction, nothing would change.  However, there are differential corrections: the V band has a larger shift than the I-band, so the differential when raised to the 4th power becomes a correction of around 10\%.

For illustration, we consider the photopolarimetric measurements of the carbon star R~Scl which exhibits comparable values for $p_\ast$ and $p_I$ from \cite{2002A&A...391..625Y}. Those authors obtained VRI polarimetry of the star. The total polarizations are around $\sim 0.5\%$, and using the catalogue of \cite{2000AJ....119..923H}, they estimate an interstellar polarization of around $0.2\%$, a significant fraction of the total polarization. The star is quite cool at around 2650~K \citep{2017A&A...601A...3W}, and it exhibits multiple types of variability \citep[see also below; e.g.,][]{2002A&A...391..625Y,2016MNRAS.463L..74K}. It also displays a number of interesting features such as extended dust shells and has been imaged with interferometry \citep{2011A&A...525A..42S}.

\cite{2002A&A...391..625Y} suggest that the polarization at optical wavelengths is roughly consistent with Rayleigh scattering involving fine dust particles. However, R~Scl varies photometrically, polarimetrically, and has been shown to exhibit variable inverse P-Cygni profiles arising in the star's warm molecular layer adjacent to the photosphere \citep{2002A&A...391..625Y,2016MNRAS.463L..74K}.
Since the interstellar polarization is constant, the variable polarization must be stellar in origin. And if it can indeed be attributed to Rayleigh scattering, then the variations should scale with $\lambda^{-4}$.  This does not depend on whether the variations are stochastic or secular. In our model, which has only Rayleigh scattering and the Serkowski Law, whatever the origin of the variations, if arising from optically thin scattering, then the amplitude of polarimetric variations should scale with the 4th power of wavelength.

\begin{table}[t]
\centering
\caption{VRI Photopolarimetry of R Scl \label{tab:rscl}}
\begin{tabular}{ccc}
\hline\hline
Filter & $p$ (\%) & $\sigma$ (\%) \\
\hline
$V$ & 0.629 & 0.106 \\
$V$ & 0.618 & 0.088 \\
$V$ & 0.509 & 0.073 \\
$V$ & 0.504 & 0.080 \\
$V$ & 0.363 & 0.091 \\
$V$ & 0.310 & 0.059 \\
$V$ & 0.265 & 0.061 \\
 & & \\
$R$ & 0.554 & 0.039 \\
$R$ & 0.503 & 0.098 \\
$R$ & 0.472 & 0.040 \\
$R$ & 0.440 & 0.035 \\
$R$ & 0.399 & 0.040 \\
$R$ & 0.359 & 0.033 \\
$R$ & 0.345 & 0.034 \\
$R$ & 0.298 & 0.040 \\
 & & \\
$I$ & 0.460 & 0.026 \\
$I$ & 0.444 & 0.047 \\
$I$ & 0.427 & 0.029 \\
$I$ & 0.394 & 0.028 \\
$I$ & 0.357 & 0.027 \\
$I$ & 0.319 & 0.079 \\
\hline
\end{tabular}
\end{table}

To explore this idea, we have taken the values of the VRI photopolarimetry of R~Scl from \cite{2002A&A...391..625Y}, listed in Table~\ref{tab:rscl}, showing the percent polarization $p$ and associated measurement uncertainty $\sigma$. For each passband we calculate the average polarization and the standard error in the mean.  We then calculate the reduced chi-squared statistic for each passband, finding values of around 2--5.  This indicates that the polarimetric variability is inconsistent with a steady source, in relation to the quoted measurement uncertainties. We then find the standard deviations of the fluctuations for each passband, which are $\sigma_V = 0.153\%$, $\sigma_R = 0.084\%$, and $\sigma_I=0.055\%$.

Qualitatively as expected from Rayleigh scattering, the fluctuations have larger amplitude at shorter wavelengths. To test quantitatively, we next scale these standard deviations by $\lambda_{\rm p}$ for each respective passband using the curves displayed in Figure~\ref{fig6} for $T_{\rm eff} = 2650$~K. Note that the normal effective wavelengths are $\lambda_{0} = 551, 662,$ and 798~nm for the VRI passbands, respectively. For a blackbody and Rayleigh scattering, the corresponding effective wavelengths are $\lambda_{\rm p} = 554, 659,$ and 793~nm. The V band effective wavelength is shifted longward whereas the R and I bands are shifted shortward.

If thin Rayleigh scattering is the dominant polarigenic opacity, then we would expect that the scaled standard deviations defined as $\tilde{\sigma}_X = \sigma_X\lambda_{\rm p}^4$ would be approximately constant for the 3 bands. Using effective wavelengths in microns, we obtain $\tilde{\sigma}$ values of 0.014, 0.016, and 0.022 for VRI, respectively. The average is 0.017 with a spread of about 20\%.  The spread in values is actually less if a $\lambda_{\rm p}^3$ scaling is used instead.  

This scaling analysis is roughly consistent with the trends expected from Rayleigh scattering. Our treatment of effective wavelengths is certainly crude: cool stars have strong absorption bands in the optical, whereas we used only a blackbody instead of the more realistic atmosphere models that could be used to calculate passband effective wavelengths. However, even this simplistic analysis reveals that the amplitude of the variations is clearly quite steep, implying a significantly large power-law trend in wavelength.

\section{Summary}\label{summ}

Polarization can be a powerful tool for inferring or constraining non-spherical geometries in spatially unresolved stars.  The challenge is that the interstellar medium imposes a polarization in addition to the intrinsic stellar polarization.  The traditional approach is to remove the ISP by considering the polarization of stars that are relatively nearby, both in direction and distance. However, such neighboring stars may not be intrinsically unpolarized, may sample sightlines with different ISP properties, or may have too few nearby sightlines to constrain the ISP for the target star.  Consequently, the attempt to remove ISP is likely to impose additional uncertainties about the intrinsic stellar polarization if not an outright systematic error.

Fortunately, the ISP is observed to generally follow the Serkowski Law, which has a particular form that peaks at optical wavelengths and declines rapidly toward the UV and NIR regimes.  Additionally, the ISP is constant with a polarization position angle that does not change.  As a result, polarimetric variability or chromatic signatures that depart from the Serkowski Law (such as a wavelength-dependent PA) are telltale signs that a star has an intrinsic polarization.

What more can be gleaned beyond this basic conclusion? The evolved late-type stars represent an important stage of stellar evolution for showing variability, having binary companions, and producing generally more powerful mass loss compared to their main-sequence progenitors. They can also exhibit polarimetric variability, thus identifying that the polarigenic opacity is important to interpreting the geometry of observations. This polarization is most likely to arise from Rayleigh scattering or dust scattering.

Here, we focused on thin circumstellar Rayleigh scattering due to its well-known wavelength dependence.  We developed a hybrid model that combines the canonical Rayleigh trend with that of the Serkowski Law and conducted a parameter study to explore spectropolarimetric effects when polarization amplitudes or PAs are varied among the stellar or interstellar contributions. We thus explored the extent to which the ISP could mask the influence of Rayleigh scattering by altering the slope of the polarization with wavelength in the optical waveband. We show that slopes for the polarized spectra are fairly steep, with trends similar to Rayleigh scattering towards shorter wavelengths. However, the polarization slopes depart from the Rayleigh variation both where the Serkowski Law tends to peak (typically around $6000~\AA$), and toward the NIR.  It would appear that Rayleigh effects could be more easily identified by sampling the shorter wavelengths. 

However, the shorter wavelengths are well on the Wien side of the flux distribution for cool stars of K, and especially M, spectral class. Even with photopolarimetry, the shorter wavelength Johnson U band can be challenging for smaller telescopes.  Even the Johnson B filter can be challenging. This generally leaves Johnson VRI filters and longer wavelength passbands, but these are exactly the wavelengths at which the Serkowski law will have its largest effects. 

Our main conclusion is that Rayleigh scattering should not be entirely discounted as the chief contributor to the polarization of cool stars, even if the polarization fails to adhere to the standard $\lambda^{-4}$ prediction. To rule out this conclusion, one would need prior knowledge that the ISP is low, or that the stellar polarization is quite large overall (which would favor the polarization as being intrinsic to the star if the source distance is not overly large, e.g., not beyond hundreds of parsecs).

We also calculated effective wavelengths associated with polarized light from Rayleigh scattering and the spectra of cool stars. This is relevant for photopolarimetry that is typically employed for optical-band polarization measurements of stars. Being on the Wien side of the flux distribution differentially shifts effective wavelengths redward of the nominal passband effective wavelengths, with more shift for filters of shorter wavelengths. However, for the polarized flux, the shift can be blueward for the long-wavelength passbands due to the $\lambda^{-4}$ scaling for Rayleigh scattering. The shifts are not large, but when testing for consistency with the expectation for Rayleigh scattering, the shifts could lead to 10\% and 20\% effects for interpreting the slope in some cases.

An underlying assumption of our model that serves as a model control is that the polarization contribution by the ISM has a fixed PA at all wavelengths. There are situations where this may not hold \citep{1974ApJ...187..461M}. If the sightline to a star transits interstellar components with different orientations of grain alignment, this can lead to a wavelength-dependent PA for the interstellar polarization \citep[c.f., the App.\ of][]{2010A&A...510A.108P}. On the other hand,
variable polarization is always stellar in origin since the interstellar contribution is constant with time. As a result, when Rayleigh scattering is the dominant polarigenic opacity for the star, the amplitude of the variation should scale with $\lambda^{-4}$ irrespective of the Serkowski parameters. We explored this concept in relation to the carbon star, R~Scl, which has previously been reported as having polarization consistent with Rayleigh scattering. It is also variable. With only three passbands and a limited sample of measurements, the variations were found to be chromatic and roughly consistent with Rayleigh scattering.

\backmatter

\begin{appendices}

%\section{Polarization Slope When Stellar and Interstellar Have the Same Position Angle}
\section{Hybrid Models with Equal Position Angles} \label{sec:app}

When $\psi_I=\psi_\ast$, we may take the position angles to be zero without loss of generality. In this case, the total polarization from combining Rayleigh scattering from the star with the Serkowski Law for the interstellar contribution becomes:

\begin{equation}
p(\lambda) = p_I(\lambda) + p_\ast (\lambda).
\end{equation}

\noindent  The result is a polarization distribution that is not a power-law, but which has a slope that varies smoothly (although not monotonically) as a function of wavelength.  As long as neither contribution is especially small, the Rayleigh influence tends to dominate at UV wavelengths, with the Serkowski effects becoming important in the NIR regime. The Rayleigh contribution could become dominant again at sufficiently long wavelength, since Serkowski has a Gaussian decline toward the IR. However, as discussed in Sect.~2.2, a power-law decline as $\lambda^{-2}$ for the interstellar polarization is better supported by observations. Since this is less steep than Rayleigh scattering, interstellar polarization will generally exceed the Rayleigh contribution.

To characterize the polarization, it is convenient to use $d\ln p/d\ln \lambda$ for the instantaneous power-law exponent, with
\begin{equation}
\frac{d\ln p}{d \ln \lambda} = \left(\frac{p_I}{p}\right)\,\frac{d\ln p_I}{d\ln \lambda}
    + \left(\frac{p_\ast}{p}\right)\,\frac{d\ln p_\ast}{d\ln \lambda}.
\end{equation}
\noindent For the stellar contribution,
\begin{equation}
\frac{d\ln p_\ast}{d\ln \lambda} = -4\, \frac{ \left( \lambda/\lambda_0 \right)^4}{1+\left( \lambda/\lambda_0 \right)^4},
\end{equation}
\noindent assuming that $\kappa^2 = 2/\lambda_0^2$ for simplicity.  For the interstellar contribution,
\begin{equation}
\frac{d\ln p_I}{d\ln \lambda} = -2\,K\,\ln\left(\lambda/\lambda_{\rm max}\right).
\end{equation}

\begin{figure}[h]%
\centering
\includegraphics[width=\columnwidth]{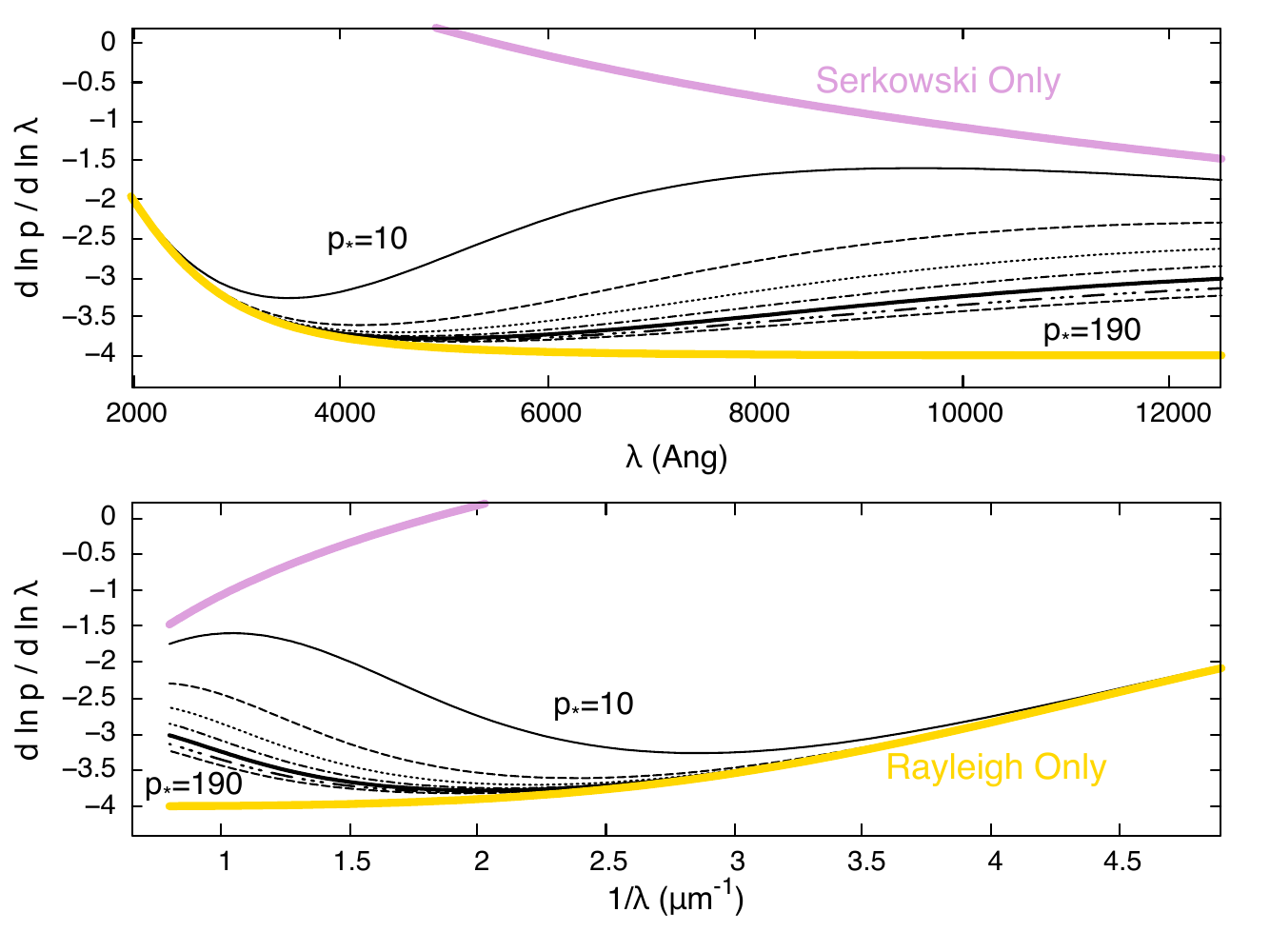}
\caption{
An example of the variation in power-law slope from the UV to the NIR regimes as expressed by $d\ln p/d\ln \lambda$.  The lower panel shows the variation in terms of inverse wavelength, with the upper panel expressed in terms of wavelength. Fixed values of $K=1.68$, $\lambda_{\rm max} = 6000~\AA$, $p_{\rm max}=0.1\%$, and $\lambda_0=2000~\AA$ are adopted.  The seven different curves are for $p_\ast = 10\%-190\%$ in steps of 30\% from top to bottom.  When the stellar polarization is stronger, the power-law slopes are closer to $-4$ in the optical because the effects of Rayleigh scattering are increasingly important.  Note that the plum colored curve is for solely the Serkowski law. Similarly, the gold colored curve is for only Rayleigh scattering.
}\label{figapp}
\end{figure}

Figure~\ref{figapp} provides an illustration, with an example of power-law slopes as a function of wavelength.  As described in the figure caption, all model parameters are fixed except for $p_0$, which sets the asymptotic polarization of the star at short wavelengths. The seven curves are for $p_0=10\%-190\%$ from top to bottom, in steps of 30\%.  For reference, $p=1\%$ at $\lambda=5000~\AA$ for $p_0=40\%$ with $\lambda_0=2000~\AA$. Even though $p_{\rm max}$ is only 0.1\% for the ISP, a power-law slope of $-4$ is never quite achieved for any of the curves, but gets close at the shortest wavelengths. How close will generally depend on the selection of $\lambda_0$.  

In this diagram, the limiting cases of only interstellar or only stellar polarization are also given as the colored upper (plum) and lower (gold) bounding curves. The plum curve traces polarization that is purely due to the Serkowski law (i.e., $p_0=0$). Different curves would result for different values of $K$ and $\lambda_{\rm max}$. The gold along the bottom derives from only Rayleigh scattering (i.e., $p_{\rm max}=0$). Again, different curves would result as a function of $\lambda_0$. Note that the bounding curve for the Serkowski law alone is independent of $p_{\rm max}$ and for Rayleigh scattering alone is independent of $p_0$.

\end{appendices}

%%===========================================================================================%%
%% If you are submitting to one of the Nature Portfolio journals, using the eJP submission   %%
%% system, please include the references within the manuscript file itself. You may do this  %%
%% by copying the reference list from your .bbl file, paste it into the main manuscript .tex %%
%% file, and delete the associated \verb+\bibliography+ commands.                            %%
%%===========================================================================================%%

\section*{Acknowledgements}

We express appreciation to an anonymous referee who's comments led
to improvements of this manuscript.

\section*{Funding Statement}

The authors declare that no funds, grants, or other support were received during the preparation of this manuscript.

\section*{Ethics Approval}

Not applicable

\bibliography{rayleigh}% common bib file
%% if required, the content of .bbl file can be included here once bbl is generated
%%\input sn-article.bbl

\end{document}